%% file: lhcp2014_kuensken.tex
\documentclass[10pt]{article}
\usepackage{graphicx}

\def\Title#1{\begin{center} {\Large #1 } \end{center}}
\def\Author#1{\begin{center}{ \sc #1} \end{center}}
\def\Address#1{\begin{center}{ \it #1} \end{center}}

\newcommand\pubblock{\rightline{\begin{tabular}{l} Proceedings of the Second Annual LHCP\\ \pubnumber\\
         \pubdate  \end{tabular}}}

\newenvironment{Abstract}{\begin{quotation} \begin{center} 
             \large ABSTRACT \end{center}\bigskip 
      \begin{center}\begin{large}}{\end{large}\end{center} \end{quotation}}

\newenvironment{Presented}{\begin{quotation} \begin{center} 
             PRESENTED AT\end{center}\bigskip 
      \begin{center}\begin{large}}{\end{large}\end{center} \end{quotation}}

\input econfmacros.tex

\usepackage{color}

\newcommand\pubnumber{ CMS CR-2014/180  }

\newcommand\pubdate{\today}

\def\affiliation{
On behalf of the CMS Collaboration, \\
Physics Institute III B\\
RWTH Aachen University, Templergraben 55, 52056 Aachen, Germany}

\begin{document}

\large
\begin{titlepage}
\pubblock

\vfill
\Title{  SiPM Operational Experience in Outer HCAL in CMS }
\vfill

\Author{ Andreas K\"unsken  }
\Address{\affiliation}
\vfill
\begin{Abstract}
The CMS Outer Hadron Calorimeter (HO) is the first large-scale hadron collider detector to use Silicon Photomultipliers (SiPMs). To build the system we installed and characterized 3000 SiPMs of which 1656 channels of SiPMs with 40MHz readout have currently been installed into CMS. We report on comparisons of in-situ and vendor-supplied measurements. We present results on working point optimization by I-V scanning and temperature vs voltage scanning. We have developed several techniques for determining the breakdown voltage in-situ. We compare the performance of each technique and its success in working point optimization. We present results on gain and breakdown voltage monitoring as well as the overall system stability.
\end{Abstract}
\vfill

\begin{Presented}
The Second Annual Conference\\
 on Large Hadron Collider Physics \\
Columbia University, New York, U.S.A \\ 
June 2-7, 2014
\end{Presented}
\vfill
\end{titlepage}
\def\thefootnote{\fnsymbol{footnote}}
\setcounter{footnote}{0}
%

\normalsize 


\section{Introduction}
\begin{figure}[htb]
\centering
\begin{minipage}[t]{0.48\textwidth}
\centering
\includegraphics[height=2in]{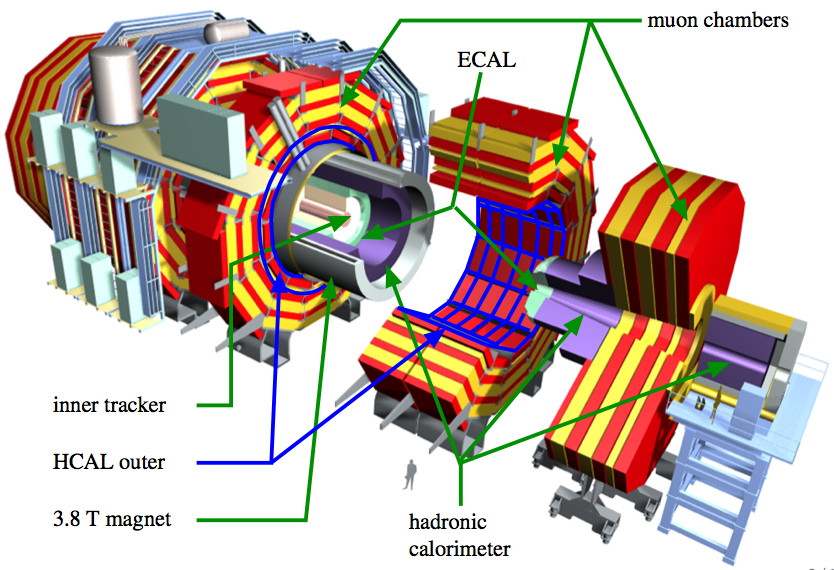}
\caption{Sketch of the CMS detector with the position of HO marked in blue \cite{lutzCalor}.}
\label{fig:cms}
\end{minipage}\hspace{1pc}
\begin{minipage}[t]{0.48\textwidth}
\centering
\includegraphics[height=2in]{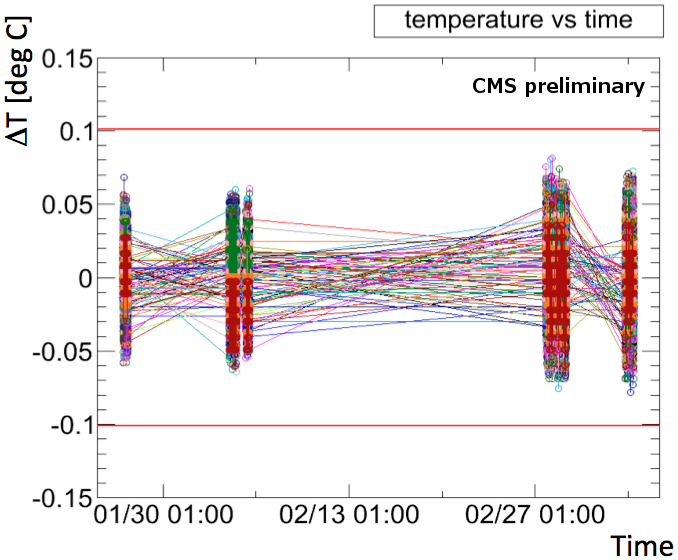}
\caption{Deviation from set temperature at the PCBs versus time \cite{kuenskenCalor}.}
\label{fig:temperature}
\end{minipage}
\end{figure}

The Outer HCAL (HO) in CMS, which is shown in fig. \ref{fig:cms} is placed behind the solenoid as a tail catcher \cite{cmsDesign}. It consists of scintillator tiles with embedded wavelength shifting (WLS) fibers in sigma shape. The fibers are coupled to clear fibers which then guide the light to the readout. While in rings $\pm\,1$ and $\pm\, 2$ there is only one layer of scintillator, ring 0 features a second layer behind the first structure of the ion flux return yoke. During the long shutdown 1 of the LHC, the hybrid photodiodes, which were used for the readout initially, were replaced with silicon photomultipliers (SiPM). 18 SiPMs corresponding to the 18 channels of one hybrid photodiode were arranged on a carrier PCB to be able to leave the rest of the readout chain untouched \cite{lutzCalor}. SiPMs are sensitive to temperature variations, therefore, there is a Peltier element for temperature stabilization on the back of each PCB.
One special feature for the readout in ring 0 arises from the fact that there are two HO layers. Thus, the double number of fibers arrives at one SiPM which do not completely fit on the SiPM's surface. To compensate for light losses at only certain fibers, a reflective light mixer is put between fiber end and SiPM. This way light losses are homogeneously distributed over the whole SiPM surface.\\
The used SiPM is a Hamamatsu Multi-Pixel Photon Counter, identical to the S10931-050P model, which has an active area of $(3\times3)\,\mbox{mm}^2$ in an SMD type housing. It has a cell pitch of $50\,\mu\mbox{m}$ and an operating voltage of the order of 70\,V. At the foreseen operating point the change of gain with temperature is approximately -8\,\%/K.

\section{Stability of temperature, gain and breakdown voltage}
\begin{figure}[htb]
\centering
\begin{minipage}[t]{0.48\textwidth}
\centering
\includegraphics[height=2in]{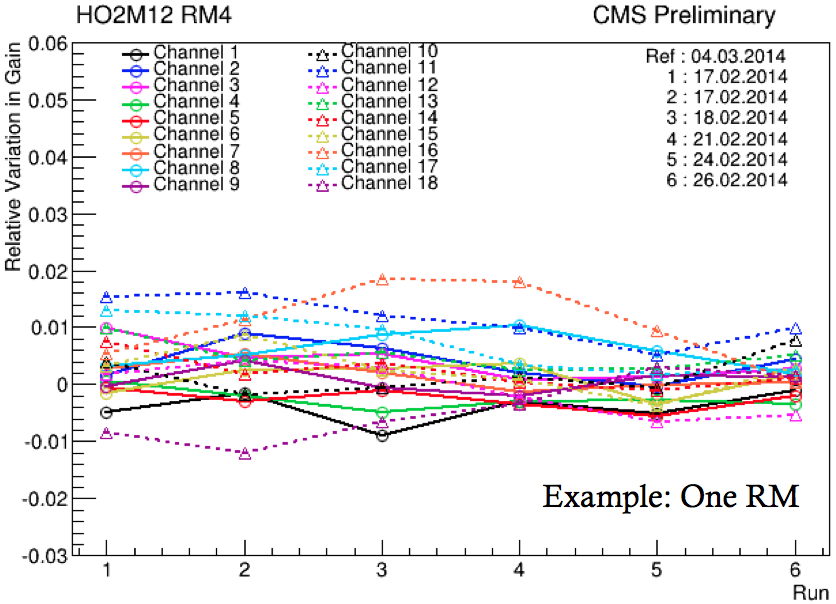}
\caption{Relative variation of gain vs. time for one PCB \cite{kuenskenCalor}.}
\label{fig:gainOverTime}
\end{minipage}\hspace{1pc}
\begin{minipage}[t]{0.48\textwidth}
\centering
\includegraphics[height=2in]{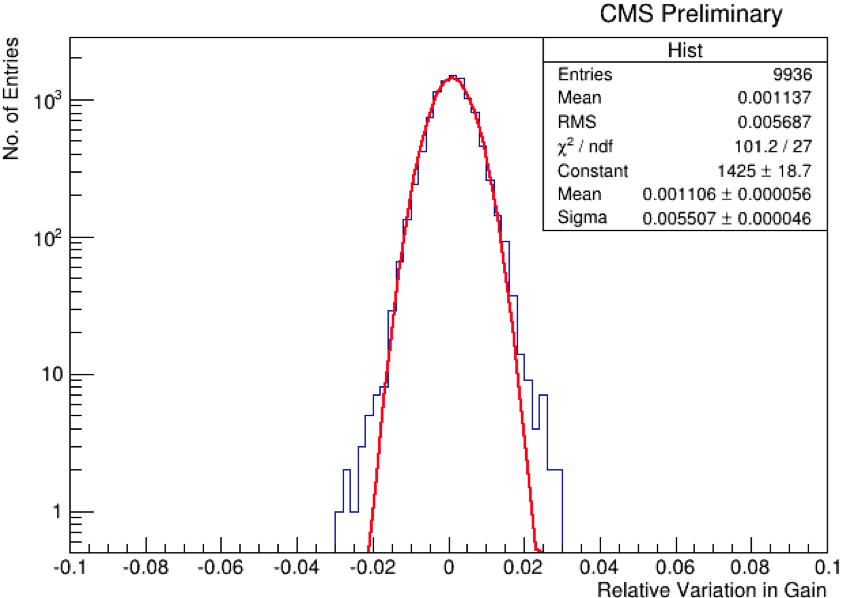}
\caption{Relative variation of gain for all PCBs from middle of February to beginning of March 2014 \cite{kuenskenCalor}.}
\label{fig:gainTotal}
\end{minipage}
\end{figure}

Figure \ref{fig:temperature} shows the temperature deviation for all installed PCBs vs. time. The measurements were taken between end of January and beginning of March 2014. The results show that the temperature environment at the PCB is stable and deviations are well within a range of $\pm$ $0.1^\circ$\,C.\\
For determining the gain there are two methods, one using the data from the dark noise spectrum of an SiPM (PED method) and the other using data taken while illuminating the SiPM with light pulses from an LED (LED method). With the PED method a fit is performed to the peaks that develop in the dark noise spectrum because of different numbers of cells breaking down in the SiPM. Assuming that $N$ photons reach the SiPM, the LED method calculates the gain from the mean $m$ of the measured signal and the width $\sigma$. Taking the mean as $N$ times gain and the uncertainty as $\sqrt{N}$ times gain one can derive
\begin{equation}
\frac{\sigma^2}{m} = \frac{N\times\mbox{gain}^2}{N\times\mbox{gain}} = \mbox{gain.}
\end{equation}
Figure \ref{fig:gainOverTime} shows the relative variation of gain versus time from middle of February to beginning of March 2014 for the SiPMs on one PCB, determined using the PED method. It can be seen that for a single PCB the relative variation in gain is already below 2\,\%. Figure \ref{fig:gainTotal} shows a histogram of all relative gain variations during this time for all installed SiPMs. The uncertainty of the distribution is at 0.6\,\% and the total distribution is contained within 3\,\%.
\\
To determine the breakdown voltage, there is also a PED method and a LED method. With the PED method the gain at different bias voltages is measured and a linear fit is performed to the resulting gain distribution. The breakdown voltage is found by extrapolating to the voltage where the gain is zero.\\
For determining the gain via the LED method, short light pulses are led onto the SiPM and the resulting signal is measured at different bias voltages. Afterwards, the relative slope d$S$/($S$d$V$) is calcualted. The breakdown voltage is defined to be the voltage at which the relative slope is at maximum.
\begin{figure}[htb]
\centering
\includegraphics[height=2in]{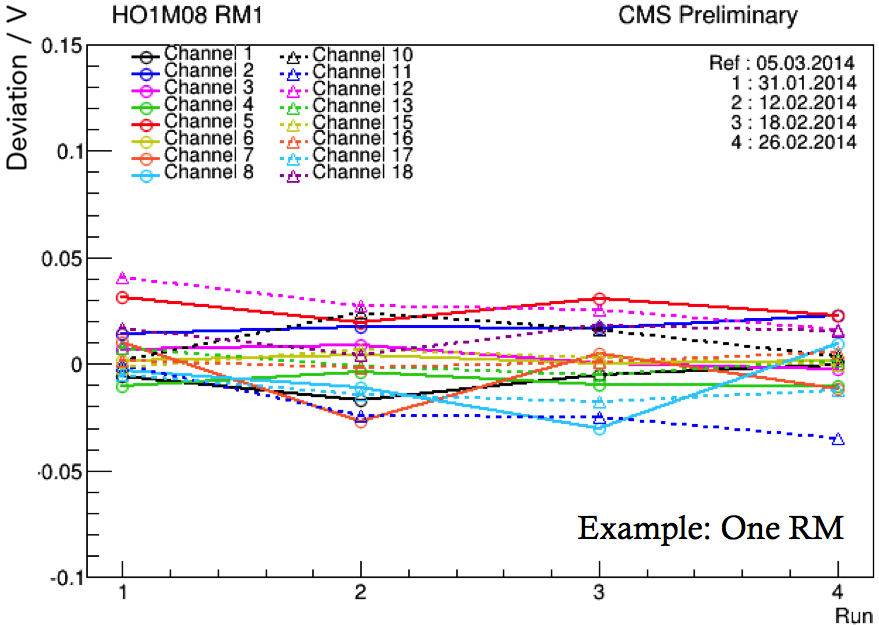}
\caption{Deviation from measured breakdown voltage over time. The results were obtained using the LED method \cite{kuenskenCalor}.}
\label{fig:bvStability}
\end{figure}
Figure \ref{fig:bvStability} displays the deviation of the breakdown voltage from the initially measured breakdown voltage for a single PCB. The data were taken between middle of February and beginning of March 2014 using the LED method. It can be seen that the breakdown voltage is stable to within 50\,mV.
\section{Correlation between methods}
\begin{figure}[htb]
\centering
\begin{minipage}[t]{0.48\textwidth}
\centering
\includegraphics[height=2in]{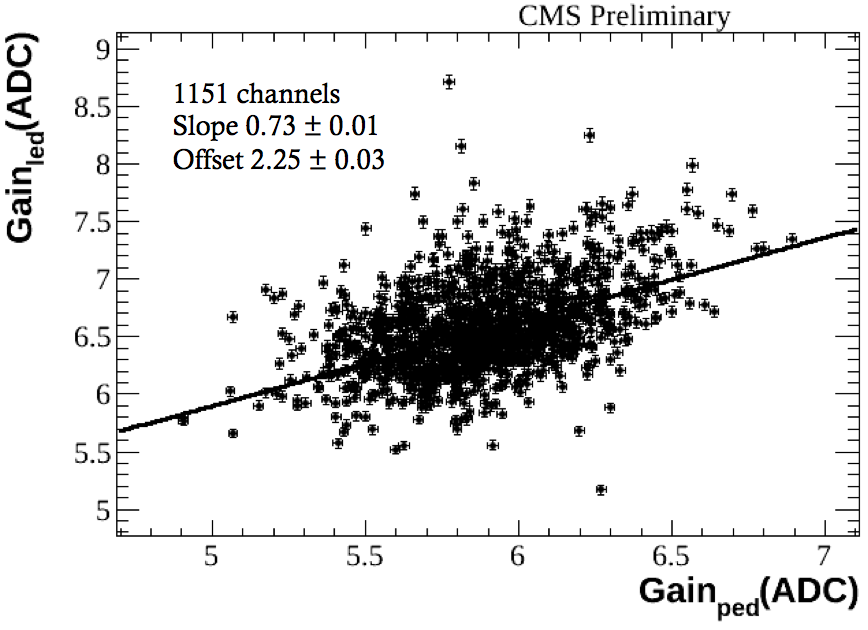}
\caption{Correlation between the LED and PED methods for gain determination \cite{kuenskenCalor}.}
\label{fig:gainCorrelation}
\end{minipage}\hspace{1pc}
\begin{minipage}[t]{0.48\textwidth}
\centering
\includegraphics[height=2in]{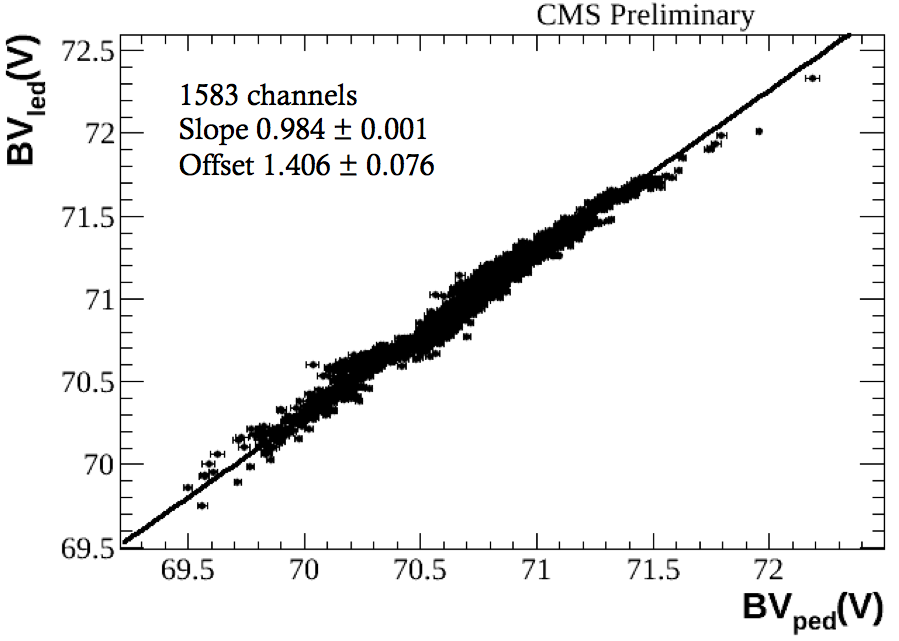}
\caption{Correlation between the LED and PED methods for breakdown voltage determination \cite{kuenskenCalor}.}
\label{fig:bvCorrelation}
\end{minipage}
\end{figure}
As there are two methods for determining the gain and the breakdown voltage, the correlation between both methods is a useful tool to spot systematic differences between them. Figure \ref{fig:gainCorrelation} shows the correlation between LED and PED method for the gain. The values are centered around a certain gain value, however, there is a systematic difference between the gain from the PED and the LED method. The results from the LED method give a higher gain. This is due to the fact that the shape of the signal distribution in the LED method is assumed to be gaussian which is not exactly the case. Once the offset between the two methods is known, it is possible to correct for it.\\
The correlation plot for the breakdown voltage is shown in Fig. \ref{fig:bvCorrelation}. The results form a line which is expected because not all SiPMs have the same breakdown voltage. Again there is an offset between the results from the two methods and the LED method gives larger results. This is explained by the fact that the definition of the breakdown voltage is different for the two methods. Having determined the offset, it is possible to correct the values from the LED method for the shift.
\section{Conclusion}
During the long shutdown 1 of the LHC HO was equipped with SiPMs to replace the HPDs. It was shown that it is possible to achieve stable operation of the SiPMs. Two methods to determine the gain and the breakdown voltage were presented and first results from the determination of gain and breakdown voltage were shown. The correlation between the PED and LED method for gain determination was studied and the found systematic difference from the LED method because of tails in the signal distribution was quantized and can be corrected for. Also the correlation of the two methods for the breakdown voltage determination was studied and the offset of the result from the LED method is characterized and can also be corrected.

\end{document}

%% file: econfmacros.tex
\def\beq{\begin{equation}}
\def\eeq#1{\label{#1}\end{equation}}
\def\eeqn{\end{equation}}

\def\beqa{\begin{eqnarray}}
\def\eeqa#1{\label{#1}\end{eqnarray}}
\def\eeqan{\end{eqnarray}}

\let\bar=\overbar

\def\etal{{\it et al.}}

\def\Dslash{\not{\hbox{\kern-4pt $D$}}}
\def\dslash{\not{\hbox{\kern-2pt $\del$}}}

\def\msb{{\bar{\ssstyle M \kern -1pt S}}}

